%% file: sample-sigconf.tex
\documentclass[sigconf]{acmart}
\newcommand{\ignore}[1]{}

\usepackage{booktabs} 
\usepackage[T1]{fontenc} 
\usepackage{enumitem}
\renewcommand\footnotetextcopyrightpermission[1]{}

\setcopyright{none}

\acmConference[GECCO '19 Companion]{Genetic and Evolutionary Computation Conference Companion}{July 13--17, 2019}{Prague, Czech Republic}

\begin{document}
\title{Toward Human-Like Summaries Generated from Heterogeneous Software Artefacts}

\author{Mahfouth Alghamdi, Christoph Treude, Markus Wagner}
\affiliation{
  \institution{School of Computer Science, The University of Adelaide, Adelaide, Australia}
  }

\begin{abstract}
Automatic text summarisation has drawn considerable interest in the field of software engineering. It can improve the efficiency of software developers, enhance the quality of products, and ensure timely delivery. In this paper, we present our initial work towards automatically generating human-like multi-document summaries from heterogeneous software artefacts. Our analysis of the text properties of 545 human-written summaries from 15 software engineering projects will ultimately guide heuristics searches in the automatic generation of human-like summaries.
\end{abstract}



\maketitle

\input{samplebody-conf}




\end{document}

%% file: samplebody-conf.tex
\begin{figure*}[ht]\vspace{-3mm}%
\centering%
\includegraphics[width=58mm]{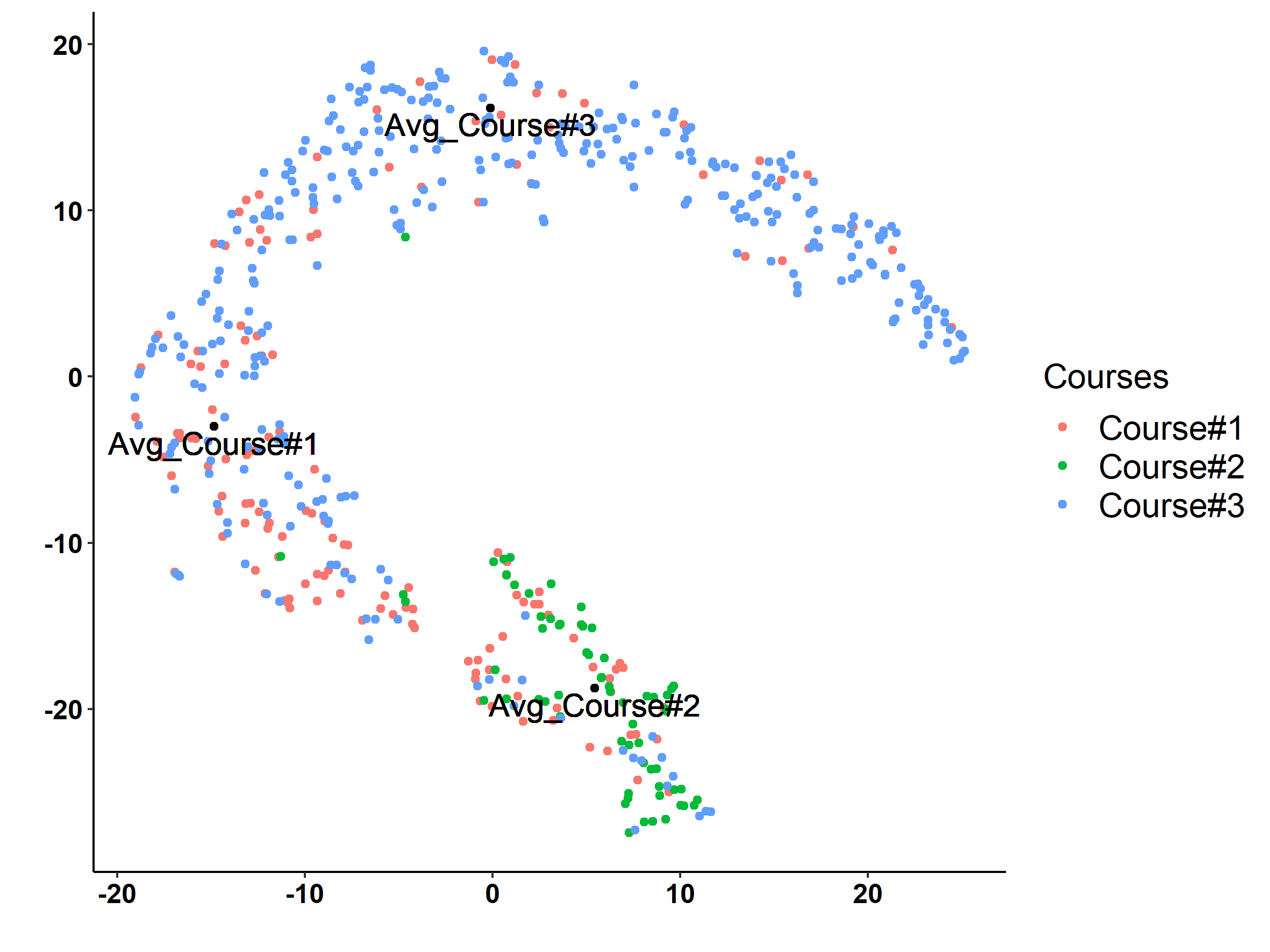}
\includegraphics[width=58mm]{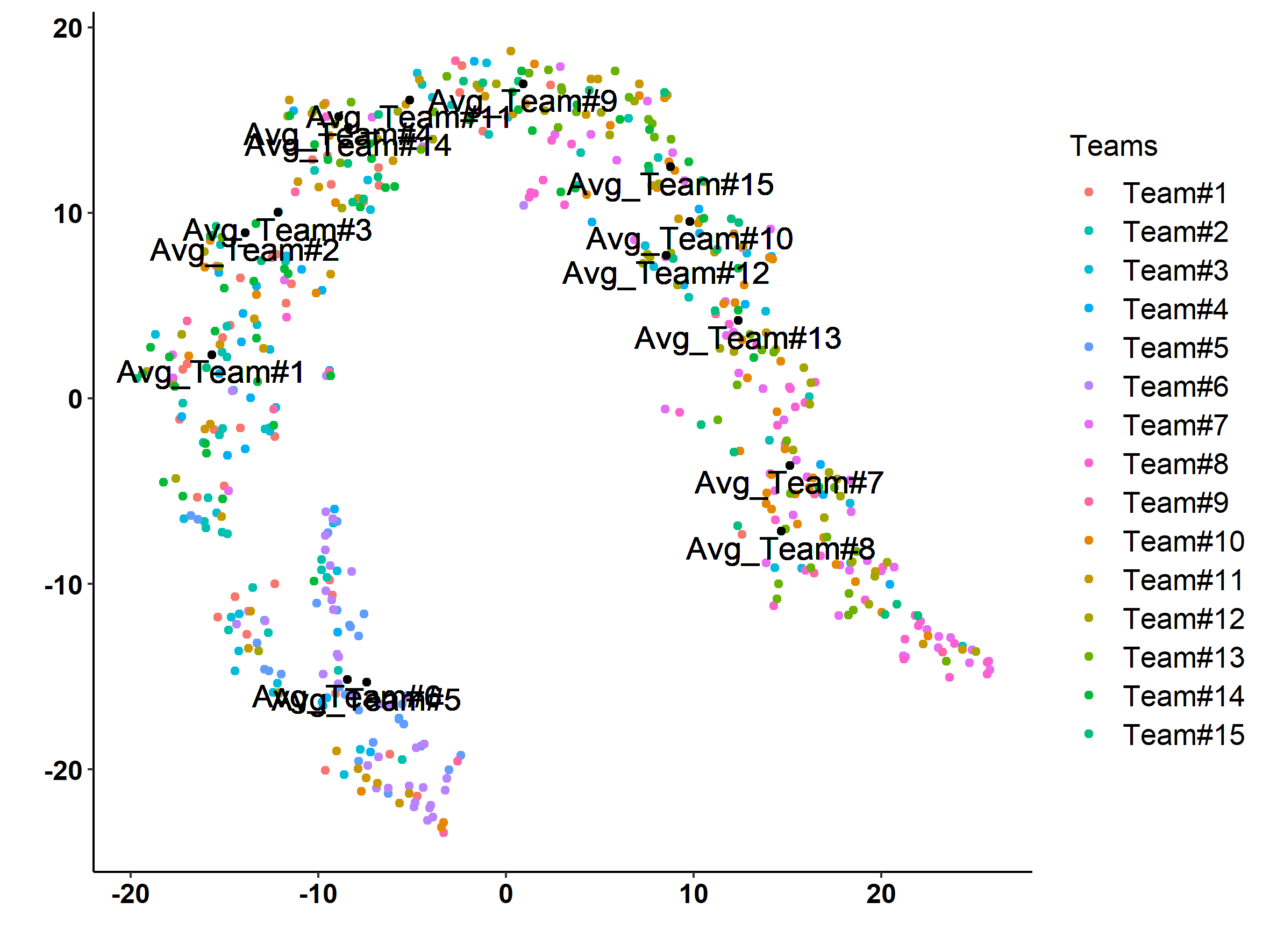}
\includegraphics[width=58mm]{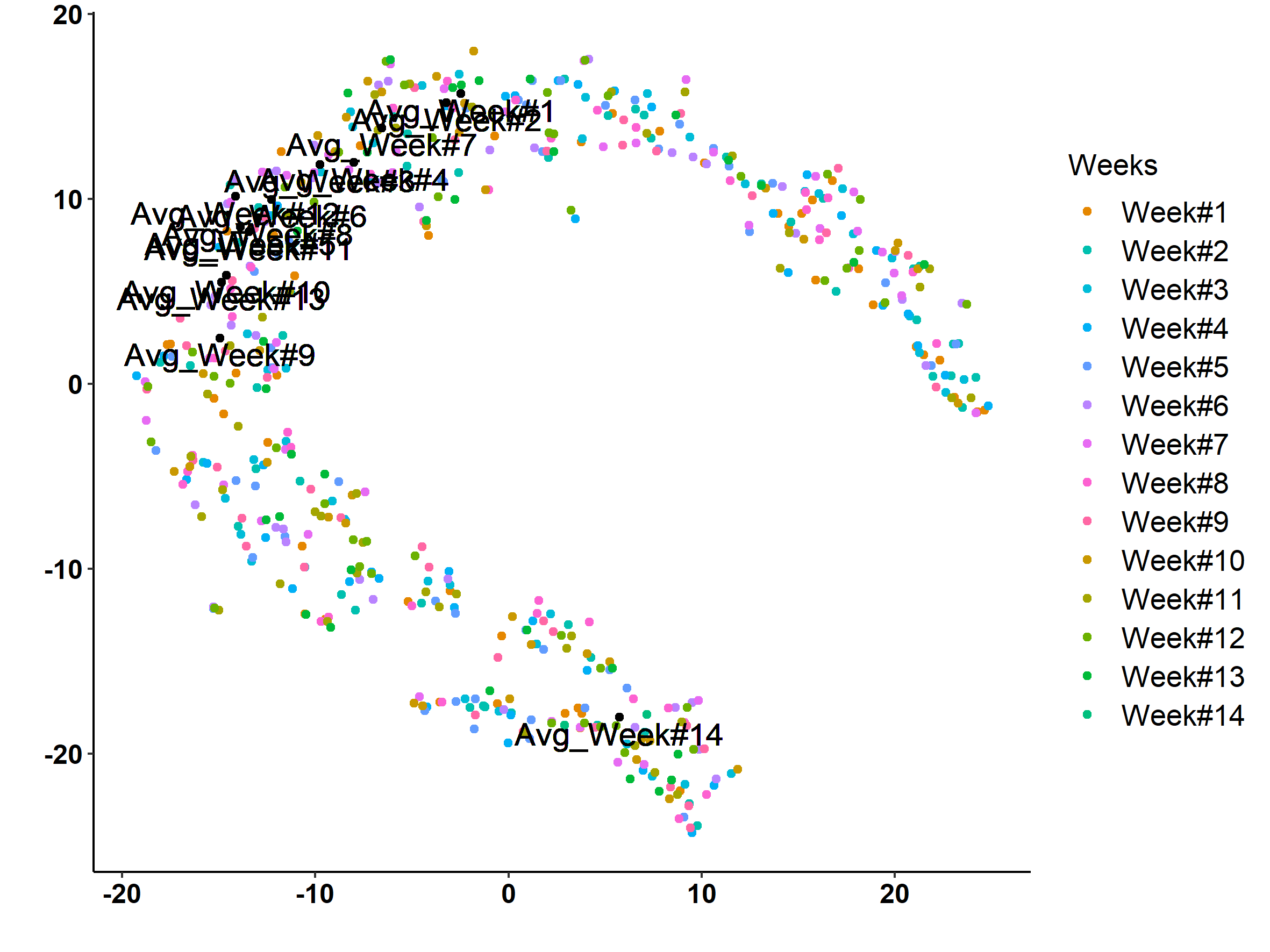}
\vspace{-3mm}%
\caption{Characteristics of the student summaries based on text features and grouped (from left to right) by courses, project teams, and weeks. The axes do not have any particular meaning in projections like these.}
\vspace{-4mm}
\label{fig:grouping-summaries}
\end{figure*}

\section{Introduction}

Since the advent of automatic text summarisation in 1958~\cite{Luhn}, summarisation techniques have been applied in various areas. In recent years, automatic text summarisation has drawn considerable interest in the area of software engineering due to the large number of software artefacts created or updated by developers. These artefacts include bug reports~\cite{Rastkar}, code elements on Stack Overflow~\cite{Rigby}, classes~\cite{Moreno}, and methods~\cite{Sridhara}.

The rise of openly available software and source code and the increase in collaborative development are facilitated by the existence of code repository services. GitHub is the leading collaborative development platform with more than 96 million repositories hosted and over 200 million pull requests, as of October 2018. There are multiple reasons for GitHub's success over other collaborative platforms. The main reason is the fact that GitHub offers more than a simple source code hosting service. It also provides developers and researchers with a dynamic and collaborative environment that supports peer reviews, commenting, and discussion~\cite{Dabbish}.

GitHub introduced a built-in search engine to allow developers to search within a project repository. This can help developers get an overview of their project activity in order to progress their work. However, the search results typically contain a large amount of heterogeneous text from many different software artefacts. For example, in the Node\footnote{\url{https://github.com/nodejs/node}, accessed on 30 March 2019.} project repository, the search results for the phrase ``test case'' contain 217 commits, 2,000 issues, and 224 source code files. This leads to several challenges in understanding developer activities regarding the search phrase. Thus, the developer is compelled to manually read through the returned artefacts to understand what has been communicated. This is extremely difficult to do when the developer is working under a limited time frame. A solution to this issue is automatic summarisation which can generate a short summary of the original text found in these heterogeneous software artefacts. 

Two main approaches have been developed for the automatic generation of summaries~\cite{Hahn}. The abstractive approach creates a summary by building a semantic representation of the source text. In contrast, the extractive approach creates a summary from a subset of existing sentences. The advantage of using the extractive techniques over the abstractive one is its capability to handle problems, such as semantic representation, natural language generation and inferences, which can be difficult for the abstractive technique to cope with~\cite{Allahyari}.

To the best of our knowledge, there is no existing approach in the context of software engineering to create multi-document summaries produced from heterogeneous software artefacts within a given time frame, yet past work has shown that developers desire such an approach~\cite{Treude2015}.
In the first step toward our ultimate goal of generating human-like summaries from heterogeneous software artefacts, we analyse a total of 545 human-written summaries produced on a weekly basis by 53 students from 15 GitHub projects to understand general properties of these summaries. The generated summaries were written in response to the question: \textit{If a team member had been away, what would they need to know about what happened this week in your project?}. A Slack bot was used to automatically ask this question on a weekly basis and to record the responses. Students were working in teams of three or four in their undergraduate capstone projects with clients from local industry toward a Bachelor degree (Courses \#1 and \#3) or with clients from academia toward their Master's degrees (Course \#2).

In the future, we plan to make use of the characterisations from this corpus of human-written summaries for our extractive summarisation approach. 
As we need to solve a subset selection problems, i.e., which sentences from which artefacts produced in a given time window should be selected, this problem can be classified as a Search-Based Software Engineering problem~\cite{harman:sbse}. 
To solve this problem, the characterisation of the students' summaries will guide the search in at least two conceivable ways:

\begin{enumerate}[noitemsep,nolistsep,leftmargin=\parindent,align=left,labelwidth=\parindent,labelsep=0pt]
\item ~~In a single-objective formulation, e.g., to minimise the cosine distance between an automatically generated summary and an ``average'' student summary.
\item ~~In a many-objective optimisation problem, where the different features are objectives in a high-dimensional space and the student summaries provide a target region to focus search on~\cite{Chand2015emo}.
\end{enumerate}

\section{Human-written summaries}

As the student summaries are supposed to guide us in the future to create human-like summaries, 
let us investigate our dataset to observe possible hidden biases and changes over time; note that both are purely observational.
First, we calculate for each of the 545 summaries 27 features related to readability metrics, lexical features, and information theoretic entropy to analyse all summaries.
Then, we visually inspect our 27-dimensional characterisation. 
To enable this, we use t-distributed Stochastic Neighbour Embedding (t-SNE)~\cite{Maaten} to project the data-points into 2D. t-SNE's reduction process attempts to preserve the distances in the high-dimensional space as much as possible.

Figure~\ref{fig:grouping-summaries} 
shows the results of grouping the summaries by weeks, courses, and teams. To facilitate the interpretation, we have added (before employing t-SNE) to each grouping the respective Euclidean average as each group's centre. Consequently, the projections unavoidably vary slightly.

\vspace{1mm}
\noindent\emph{Insights per grouping:} In the following, we highlight a few interesting observations. 
The summaries of the students grouped by courses are shown on the left in Figure~\ref{fig:grouping-summaries}. These courses are taught to graduate students (Course \#2) and undergraduate students (Course \#1 and Course \#3). Also, the student projects involved in these courses are categorised as industrial projects (Course \#1 and Course \#3) and non-industrial projects (Course \#2). It is apparent from the distribution of the summaries that Course \#2 sits apart at the bottom and far from other groups while the two other courses are close to each other at the top. The distribution of these courses reveals that the summaries generated by the graduate students whose projects are categorised as non-industrial projects have different text properties while the two other courses have similar text properties, such as length of the summaries, readabilities metrics, and entropy. This variation in the summary properties can be attributed to many factors, including type of project (industrial/non-industrial projects), education level (undergraduate/graduate students), and writing style (students in Course\#2 are less likely to be native speakers). 

Similarly, in the middle of Figure~\ref{fig:grouping-summaries}, the summaries are grouped by teams from each course. Summaries produced by Teams \#5 and \#6 have similar text properties. In the same manner, Teams \#4, \#11, and \#14 have similar text properties, but they belong to two different courses (Course \#1 and Course \#3, respectively---although these courses are different instances of the same course offered in different years). Teams \#5 and \#8 have the highest and lowest average values respectively across all teams in term of some of the features calculated. By inspecting the members of Team \#5, we found that most of the team are non-native speakers unlike Team \#8, and thus features calculated such as word count, average sentence length, and unique words are less compared to features calculated for members of Team \#8. 

Summaries written in later parts of the semester appear to have fewer development activities compared to the initial weeks where the students have much work to do to develop their projects. This is reflected in the features calculated from their summaries and shown on the right in Figure~\ref{fig:grouping-summaries}. Besides, different summaries formed by the students who belong to different teams seem to have similar text properties in different weeks. For example, Weeks \#1 and \#2 have text properties that are very close to each other. We note the same behaviour in Weeks \#10 and \#13.

\vspace{1.0mm}
\noindent\emph{Conclusion:} Our approach of utilising t-SNE to interpret the students' summaries data at different grouping levels using the 27 features  allows us to identify summaries that can serve as ``gold standard'' summaries. We will use these to evaluate our future work on extractive summarisation techniques.
\vspace{-1.6mm}